\documentclass[twocolumn, pra, showpacs,superscriptaddress]{revtex4}
\usepackage{graphicx}
\usepackage{amsmath}
\usepackage{bm}
\usepackage{float}
\usepackage{color}
\usepackage{sidecap}

\begin{document}
\title{Emergent patterns in a spin-orbit coupled spin-2 Bose-Einstein condensate}
\author{Z. F. Xu}
\affiliation{Department of Physics, Tsinghua University, Beijing
100084, People's Republic of China}
\author{R. L\"u}
\affiliation{Department of Physics, Tsinghua University, Beijing
100084, People's Republic of China}
\author{L. You}
\affiliation{Department of Physics, Tsinghua University, Beijing
100084, People's Republic of China}

\date{\today}

\begin{abstract}
The ground-state phases of a spin-orbit (SO) coupled atomic spin-2
Bose-Einstein condensate (BEC) are studied. Interesting density patterns
spontaneously formed are widespread due to the competition
between SO coupling and spin-dependent interactions like in a
SO coupled spin-1 condensate.
Unlike the case of spin-1 condensates, which are characterized by
either ferromagnetic or polar phase in the absence of SO,
spin-2 condensates can take a cyclic phase, where we find
the patterns formed due to SO are square or triangular in
their spin component densities for axial symmetric SO interaction.
Both patterns are found to continuously evolve into striped forms with
increased asymmetry of the SO coupling.
\end{abstract}


\maketitle

\section{Introduction} 

The technique of gauge field is a useful tool in theoretical
physics. Depending on commutation relations of the associated
operators, gauge fields are classified into Abelian and non-Abelian
ones. An important example for Abelian gauge field concerns the well
studied vector and scalar potentials of electromagnetic fields.
Interesting phenomena, such as integer and fractional quantum hall
effects are observed in two-dimensional high mobility electrons
inside a perpendicular magnetic field. For a collection of cold
atoms, Abelian gauge field can be affected through rotation
\cite{fetter2009, cooper2008} or adiabatic motion inside far
off-resonant laser fields
\cite{dalibard2010,juzeliunas2005,gunter2009,lin2009}. While Abelian
gauge fields are often studied, recent interests in Bose condensed
atoms are increasingly targeted at situations when gauge fields
become non-Abelian
\cite{dalibard2010,ruseckas2005,osterloh2005,satija2006}. One of the
simplest examples concerns spin-orbit (SO) coupling, which recent
understandings increasingly view as exists intrinsically in
solid-state systems \cite{zutic2004,ruostekoski2002}, and is
responsible for the quantum spin hall effect. In atomic quantum
gases, SO coupling has been proposed with a variety of approaches
involving atomic interactions with electromagnetic fields
\cite{dalibard2010,ruseckas2005,stanescu2008,wang2010,ho2010,yip2010,lin2011,merkl2010,braun2010},
including schemes for two component or pseudo spin-1/2 atomic
condensates \cite{ho2010,yip2010} and spin-1 condensates
\cite{ruseckas2005,wang2010}, and realized experimentally for pseudo
spin-1/2 system recently \cite{lin2011}.

Spin-orbit interaction couples the internal (spin)
and orbital (momentum or angular momentum)
degrees of freedom. In cold atoms, elastic binary collisions are described
by contact interactions proportional to their respective s-wave scattering lengths.
For atoms with internal degrees of freedom, their
collision interactions generally contain spin dependence,
as in spinor atom condensates \cite{ueda2010}, where the
spin-dependent interactions determine the various ground
state phases: ferromagnetic or polar for the case of spin-1;
ferromagnetic, polar, or cyclic, for spin-2; {\it etc}.
The inclusion of SO coupling induces competition with
spin-dependent interactions in addition to modify
 the single particle spectra for a spinor BEC.
As a result, spin-dependent interaction will in turn influence atomic
spatial motion, leading to a variety of density patterns
even in the ground state \cite{wang2010,ho2010,yip2010,lin2011}.

Spontaneously formed patterns in spinor component densities,
are essentially spin textures, previously studied in
$^3$He superfluids and more recently in spinor condensates
for both ferromagnetic and anti-ferromagnetic (polar) phases,
especially with long-range dipolar interactions
\cite{zhang2005,sadler2006,saito2009,cherng2009,matuszewski2010,kronjager2010,scherer2010}.
Periodically ordered patterns can also arise in superfluids
with roton-like spectra \cite{cherng2009,matuszewski2010}.

This study addresses ground state pattern formations due to
the competition between SO coupling and spin-dependent interactions.
The simplest case concerns a spin-1/2 BEC with SO coupling,
\cite{wang2010,ho2010,yip2010}, already observed in recent experiments \cite{lin2011}.
Depending on the sign of the effective spin interaction strength,
which is proportional to the difference between intra- and
inter-species atomic scattering strengths, the ground state
patterns are found to be either planar or standing waves \cite{wang2010,ho2010,yip2010,lin2011}.
The same conclusion was reached in recent theoretical studies
for a spin-1 BEC \cite{wang2010}, where the two ground state phases
are ferromagnetic and polar in the absence of SO coupling.
Spin-2 condensates, on the other hand,
can potentially be very different due to the
presence of a cyclic phase when SO coupling is absent.
Understanding the associated patterns in a spin-2 BEC thus
constitutes an importance objective for new physics in
atomic quantum gases with gauge fields.

\section{Our model}

We start by describing our model of a spin-2 BEC \cite{ciobanu2000,koashi2000}
in a quasi-two dimensional optical trap $V_o=m\omega^2
(x^2+y^2+\lambda^2z^2)/2$ with $\lambda\gg1$, including SO coupling.
The effective Hamiltonian takes the form
\begin{eqnarray}
  \hat{H}&=&\int d\bm{\rho}\,\hat{\psi}_i^{\dag}\left(\frac{\mathbf{p}^2}{2m}
  +v_xp_xF_x+v_yp_yF_y+V'_o \right)_{ij}\hat{\psi}_j\nonumber\\
  &+&\frac{1}{2}\int d\bm {\rho}\,\left\{\alpha'\hat{\psi}^{\dag}_i\hat{\psi}^{\dag}_j
  \hat{\psi}_j\hat{\psi}_i+\beta'\hat{\psi}^{\dag}_i\hat{\psi}^{\dag}_k
  \mathbf{F}_{ij}\cdot\mathbf{F}_{kl}\hat{\psi}_l\hat{\psi}_j\right.
  \nonumber\\
  &&\qquad\quad\quad\left.+\gamma'(-1)^{i+j}\hat{\psi}_i^{\dag}\hat{\psi}_{-i}^{\dag}\hat{\psi}_j
  \hat{\psi}_{-j}\right\},
  \label{hamiltonian}
\end{eqnarray}
where $(F_{\mu})_{ij}$ ($\mu=x,y,z$) are the $(i,j)$ components of
spin-2 matrices $F_{\mu}$, $V'_o=m\omega^2(x^2+y^2)/2$, $\alpha'$,
$\beta'$ and $\gamma'$ are reduced two-dimensional effective
density-density, spin-exchange, and spin singlet-pairing interaction
parameters, respectively. $v_x$ and $v_y$ parameterizes the strength
of SO coupling. This form of two- and three-component of condensates
with SO coupling can be realized with a tripod and a tetrapod
plane-wave laser beams setup \cite{ruseckas2005}, respectively.
Also, recently Y. -J. Lin {\it et al.} \cite{lin2011} have realized
a mixed form of Rashba and Dresselhaus type SO coupling interaction
for pseudo spin-1/2 system. They further suggested that we can 
realize the SO coupled spin-1 or spin-2 condensates
by using a smaller quadratic Zeeman shift.

The single particle Hamiltonian
\begin{eqnarray}
  \hat{H}_0=\frac{\mathbf{p}^2}{2m} +v_xp_xF_x+v_yp_yF_y,
  \label{hamiltonian0},
\end{eqnarray}
ignoring the external trapping potential,
can be easily diagonalized using plane wave basis,
leading to eigenvalues $E_{\lambda}(\mathbf{p})$ and eigenvectors
$\phi_{\lambda}(\mathbf{k})=e^{i\mathbf{k}\cdot\bm{\rho}}\zeta_{\lambda}(\mathbf{k})$
given by
\begin{eqnarray}
  E_{\lambda}(\mathbf{k})&=&\frac{\hbar^2\mathbf{k}^2}{2m}+\lambda\sqrt{\hbar^2v_x^2k_x^2+\hbar^2v_y^2k_y^2},
  \label{eigenvalues}
  \\
  \zeta_{2}^T(\mathbf{k})&=&(\chi_{\mathbf{k}}^{*2},2\chi_{\mathbf{k}}^*,\sqrt{6},2\chi_{\mathbf{k}},
  \chi_{\mathbf{k}}^2) /4,\nonumber\\
  \zeta_{1}^T(\mathbf{k})&=&(-\chi_{\mathbf{k}}^{*2},-\chi^*_{\mathbf{k}},0,\chi_{\mathbf{k}},\chi_{\mathbf{k}}^2)/2,
  \nonumber\\
  \zeta_0^T(\mathbf{k})&=&(\sqrt{3}\chi_{\mathbf{k}}^{*2},0,-\sqrt{2},0,\sqrt{3}\chi_{\mathbf{k}}^2)/2\sqrt{2},
  \nonumber\\
  \zeta_{-1}^T(\mathbf{k})&=&(-\chi_{\mathbf{k}}^{*2},\chi^*_{\mathbf{k}},0,-\chi_{\mathbf{k}},\chi_{\mathbf{k}}^2)/2,
  \nonumber\\
  \zeta_{-2}^T(\mathbf{k})&=&(\chi_{\mathbf{k}}^{*2},-2\chi_{\mathbf{k}}^*,\sqrt{6},-2\chi_{\mathbf{k}},
  \chi_{\mathbf{k}}^2) /4.
  \label{eigenvectors}
\end{eqnarray}
In the above $\lambda=\pm2,\pm1,0$ labels the respective energy band
and $\chi_{\mathbf{k}}=(v_xk_x+iv_yk_y)/\sqrt{v_x^2k_x^2+v_y^2k_y^2}$.
Corresponding to the above eigenvectors Eq.~(\ref{eigenvectors}),
we find the two order parameters
 $(|\langle \mathbf{F}\rangle|=|\zeta_{\lambda}^{\dag}\mathbf{F}\zeta_{\lambda}|$
 and $|\langle\Theta\rangle|=|(-)^{j}\zeta_{\lambda,j}\zeta_{\lambda,-j})|)$
are equal to $(2,0)$, $(1,0)$, and $(0,1)$ corresponding to $\lambda=\pm2,\ \pm1,\ 0$,
respectively.

The ground state of the single particle Hamiltonian
(\ref{hamiltonian0}): $\phi_{-2}(\mathbf{k}_g)$, is two-fold degenerate
for $|v_x|\ne |v_y|$; and is infinitely degenerate if $|v_x|=|v_y|$.
For the former case, $\hbar\mathbf{k}_g=(\pm 2m|v_x|,0)$ if $|v_x|>|v_y|$,
while $\hbar\mathbf{k}_g=2m|v_x|(\cos\theta_g,\sin\theta_g)$ for any
$\theta_g\in[0,2\pi)$, in the latter case with $|v_x|=|v_y|$.
Atomic spins are fully polarized with $|\langle F\rangle|=2$ for both cases.

\section{Ground-state patterns}

When SO coupling is absent, the ground states for a spin-2 BEC
including collision interactions, takes three phases: ferromagnetic,
polar, and cyclic \cite{ciobanu2000,koashi2000}.
Their corresponding spin-dependent interactions will in this study
be called ``ferromagnetic'', ``polar'', and ``cyclic'' interactions.
In the presence of SO coupling, we discuss their possible ground states
for the above three types of spin-dependent interactions.
The external trapping potential is neglected in analytical treatments,
while it is weakened appreciabely for numerical studies
with imaginary-time propagation of coupled Gross-Pitaevskii
equations (GPEs). Including axisymmetric SO interaction,
the ground states can be approximately understood by applying perturbation
theory to the degenerate single particle states.
Each of the three types of spin-dependent interactions is found
to be associated with a distinct number single-particle states.
For the case of ``cyclic'' interaction, this approximation
breaks with increasing asymmetry of SO coupling.

{\it (A) ``Ferromagnetic'' interaction}\hskip 12pt
The ferromagnetic spin-exchange interaction tends to polarize
atoms' spin, consistent with the ground state of the single-particle
Hamiltonian. As a result, all atoms condense into any single particle
ground state $\phi_{-2}(\mathbf{k}_g)$.
The spin rotation and time-reversal symmetry are broken in this case.
The optical trap and the spin-independent interactions will
change atomic density distributions slightly.

\begin{figure}[H]
\centering
\includegraphics[width=3.0in]{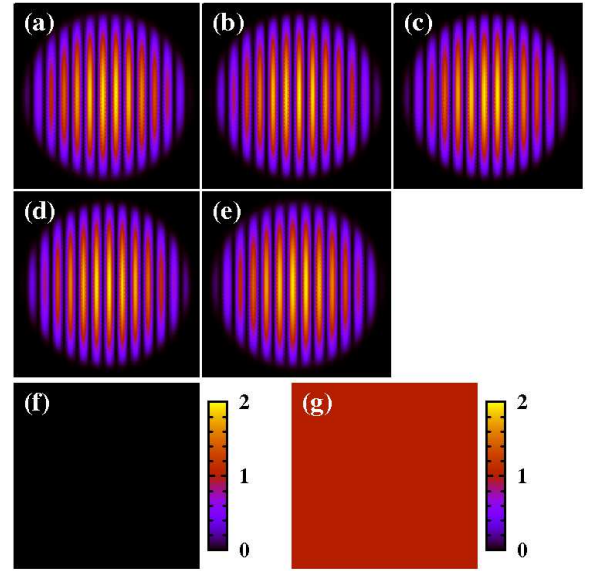}
\caption{(Color online). Ground states of a spin-2 BEC
in the ``polar'' interaction case with SO coupling for
$v_x=2v_y$ and $\gamma'=-\beta'=-0.2\alpha'<0$. (a-e) Density distributions
for the five spin components: $\rm M_F=+2,+1,0,-1,-2$.
Yellow and black colors represent high and low density respectively.
(f-g) Spatial distributions of two order parameters (uniform in this case) 
$|\langle\mathbf{F}\rangle|$ and $|\langle\Theta\rangle|$, which 
approximately measure spin-exchange and singlet-pairing interactions, respectively.}
\label{fig1}
\end{figure}

{\it (B) ``Polar'' interaction}\hskip 12pt  The order parameter $|\langle F\rangle|=2$ in the single
particle ground state conflicts with the polar phase of $|\langle F\rangle|=0$.
A superposition of degenerate single particle states is adopted
\begin{eqnarray}
  \psi=\sum_{\mathbf{k}_g}\mathcal{C}_g\phi_{-2}(\mathbf{k}_g),
  \label{wavesummation}
\end{eqnarray}
to minimize the spin-dependent interaction.
The states $\phi_{-2}(\mathbf{k}_g)$
are orthogonal, thus the superposition state $|\psi\rangle$ does not affect kinetic energy.
The suitable coefficients $\mathcal{C}_g$ should give
$|\langle\psi|\mathbf{F}|\psi\rangle|=0$ and $|\langle\psi|\Theta|\psi\rangle|=1$,
while preserving the spin-independent interaction energy.
It is easy to conclude a superposition of two counter-propagating
single particle states meets this criterion, thus the ground state
is given by $\phi_{-2}(\mathbf{k}_g)+e^{i\vartheta}\phi_{-2}(-\mathbf{k}_g)$,
where the time-reversal symmetry remains valid.
Back into the $z$-axis quantized representation, the interference between
the two counter-propagating plane waves induce density oscillations
in spinor components. In the quantization direction
of $\mathbf{k}_g$, however,
only $M_F=\pm2$ components are found to be populated,
each with smooth density, respectively propagating into opposite directions.

In Figure~\ref{fig1}, we illustrate density distributions of the spin
components and the spatial dependence of two order parameters
$|\langle\mathbf{F}\rangle|$ and $|\langle\Theta\rangle|$
for $v_x=2v_y$ and $\gamma'=-\beta'=-0.2\alpha'<0$.
When $|v_x|>|v_y|$ ($|v_x|<|v_y|$), we find
the plane waves propagate along the positive/minus $x$-axis ($y$-axis)
the ground state, resulting in density modulations along $x$-axis ($y$-axis).
Figure~\ref{fig1}(f-g) shows the two order parameters are uniformly distributed,
with $|\langle\mathbf{F}\rangle|=0$ and $|\langle\Theta\rangle|=1$, consistent with
previous discussions.
A cautionary note concerns the axis-symmetric case of $|v_x|=|v_y|$.
Due to the spin-independent interaction, the ground state consists of only
one pair of counter-propagating plane waves, although
superpositions of two or more counter-propagating pairs also minimize
spin-dependent energy.

\begin{figure}[tpb]
\centering
\includegraphics[width=3.2in]{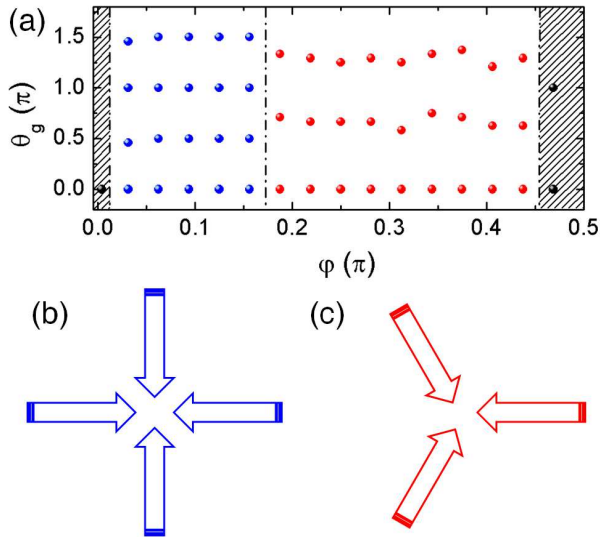}
\caption{(Color online). Ground states of a spin-2 BEC
in the ``cyclic'' interaction case with SO coupling for $v_x=v_y$.
(a) The number of plane waves in the ground states
with changing $\varphi={\rm arg}(\gamma'+i\beta')$.
The column of 4 blue (3 red) dots denote four (three) plane waves.
The shadow window with one or two black dots denote one or two
plan waves respectively constituting of the ground states.
Schematic illustrations of four b) [three c)] plane wave superpositions
responsible for square (triangular) density modulations.
The angles between the directions of two nearby plane waves are close
but not exactly equal to $2\pi/4$ or $2\pi/3$ due to frustration.
}
\label{fig2}
\end{figure}

\begin{figure}[tpb]
\centering
\includegraphics[width=3.0in]{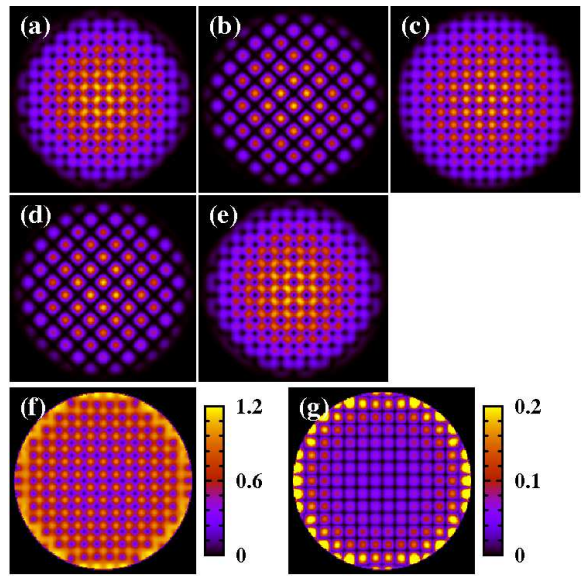}
\caption{(Color online). Ground states of a spin-2 BEC
in the ``cyclic'' interaction case with SO coupling for
$v_x=v_y$, $\gamma'=0.2\alpha'$, and $\beta'=0.02\alpha'$.
(a-e) Density distributions of the
spin components: $\rm M_F=+2,+1,0,-1,-2$. Yellow and black colors
represent high and low densities respectively.
(f-g) Spatial distributions of the order parameters
$|\langle\mathbf{F}\rangle|$ and $|\langle\Theta\rangle|$, reflecting
spin-exchange and singlet-pairing interactions respectively.}
\label{fig3}
\end{figure}

\begin{figure}[tpb]
\centering
\includegraphics[width=3.0in]{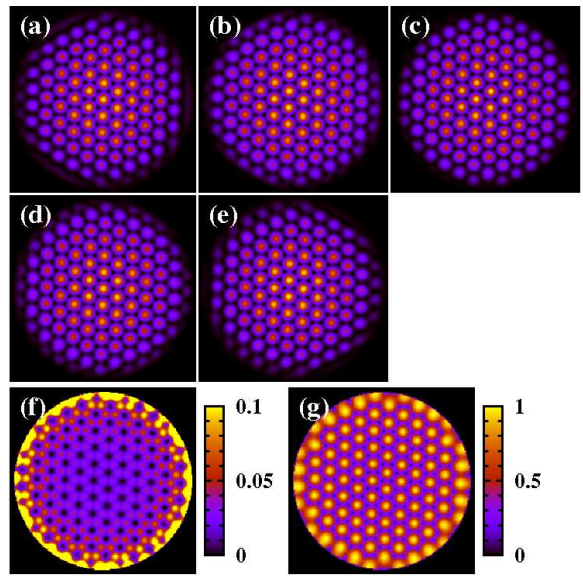}
\caption{(Color online). The same as in Fig. \ref{fig3}
but for $v_x=v_y$, $\gamma'=0.02\alpha'$ and $\beta'=0.2\alpha'$.}
\label{fig4}
\end{figure}

\begin{SCfigure*}
\centering
\includegraphics[width=0.6\textwidth]{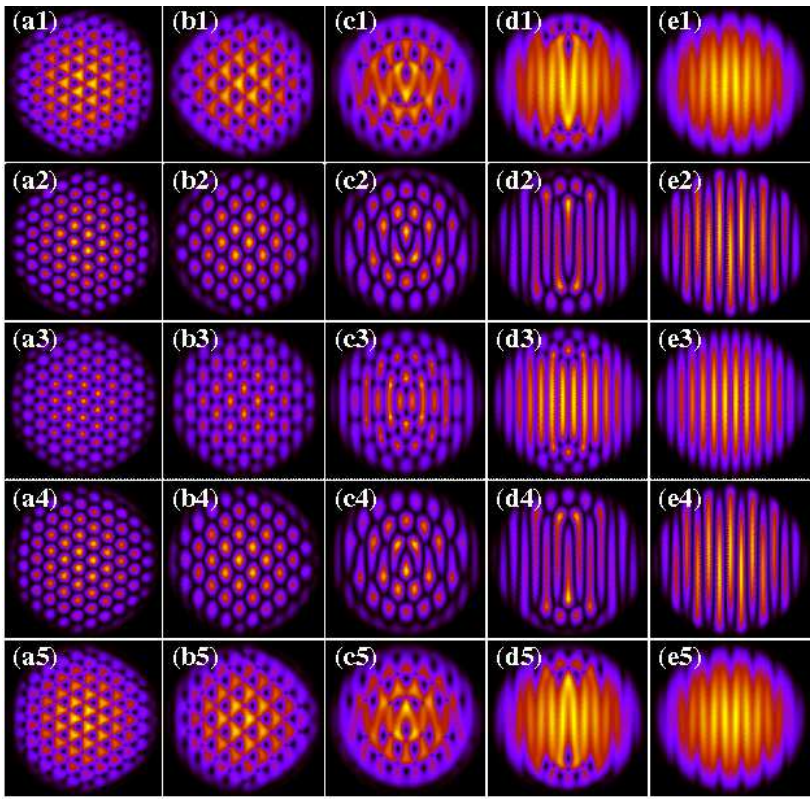}
\caption{(Color online). Ground states of a spin-2 BEC
in the ``cyclic'' interaction case with SO coupling for
$\gamma'=\beta'=0.2\alpha'$. (a1-a5) Spinor component densities
at $v_y=v_x$, from up to down corresponding to
$M_F=+2,+1,0,1,2$. (b1-b5), (c1-c5), (d1-d5), and (e1-e5) are
as in (a1-a5) but for $v_y=3v_x/4$, $v_y=17v_x/24$, $v_y=2v_x/3$,
and $v_y=v_x/2$,
respectively.}
\label{fig5}
\end{SCfigure*}

{\it (C) ``Cyclic'' interaction}\hskip 12pt First, we consider the
symmetric case of $|v_x|=|v_y|$. As pointed out in the above {\it
``polar'' interaction} case, superpositions of single particle
ground states are tried to minimize the interaction energy.
Numerically, the minimization procedure is achieved with simulated
annealing. In Fig.~\ref{fig2}(a), the ground states are found to
contain several plane waves with significant weights if they are
assumed approximately to consist of a series of degenerate single
particle ground states with momentum
$\hbar\mathbf{k}_g=2m|v_x|(\cos\theta_g,\sin\theta_g)$ and if
external optical trapping potentials are neglected. In this case,
the spin-dependent interactions couple degenerate single particle
ground states, and result in discrete rotation symmetry of the
ground states. Starting from the boundary of $\beta'=0$, with
increasing value of $\varphi={\rm arg}(\gamma'+i\beta')$, the ground
states at first contains only one plan wave, as $\varphi$ exceeds a
small critical value determined by the three interaction parameters,
the ground states display periodic square lattices in the spin
component densities. Further increasing of $\varphi$ arrives at the
critical value $\varphi_c$, where the ground states change into
triangular from superpositions of three plane waves until near the
boundary with $\gamma'=0$, where the ground state superpositions
reduce to contain two counter-propagating plan waves. It is
important to emphasize that interference between serval plane waves
may increase density-density interaction energy, which explains why
the phase boundaries between the ground states with four and one,
also with three and two plan waves, deviate from $\varphi=0$ or
$\varphi=\pi/2$. The numerical results show for the two different
types of ground states with square or triangular patterns, the
angles between nearby plane waves are close to but not exactly equal
to $2\pi/4$ or $2\pi/3$ respectively, and the nonzero coefficients
$\mathcal{C}_g$ are not all exactly equal, although they are of the
same orders of magnitudes.

To validate our proposition that the ground states of
a spin-2 BEC with SO interaction are approximate superpositions
of degenerate single particle ground states in the absence of
the trap potential, we numerically solve the GPEs directly using
imaginary-time propagation including the optical trap.
Figures~\ref{fig3} and \ref{fig4} show spin component
densities and the spatial dependence of the
two order parameters $|\langle\mathbf{F}\rangle|$
and $|\langle\Theta\rangle|$ for two different cases.
They confirm our understanding that not only spin component density distributions
but the two order parameters show square or triangular patterns.
More carefully, the shapes of density distributions within each unit cell
are not identical for the five spin components at
a fixed $\varphi$. Their differences vary as
$\varphi$ is changed. Because the values of
$\mathcal{C}_g$ are not exactly equal, they generally vary
with $\varphi$ constrained by the spin-independent interaction
term, although the number of plane waves
contained in the ground states remains at four or three.

Next, we consider the asymmetric case of $|v_x|\ne|v_y$, where the
single particle ground states are two-fold degenerate. When
insufficient number of degenerate single particle ground states
exist to construct square or triangular shaped density
distributions, our proposition fails. This is also validated by
numerical simulations. In Fig. \ref{fig5}, we illustrate results
from numerical simulations. From (a1-a5), (b1-b5), (c1-c5), (d1-d5),
to (e1-e5), the asymmetry of the SO coupling is increased, realized
by fixing $v_x$ while decreasing $v_y$ from being equal $v_x$,
$3v_x/4$, $17v_x/24$, $2v_x/3$ to $v_x/2$, respectively. The ground
state show triangular density distributions in each component at
$v_y=v_x$. The triangular lattice is deformed into almost square
lattice distributions when $v_y$ is decreased into $3v_x/4$, and
gradually evolve in the end into striped patterns reflecting the
underline two-fold degeneracy as we further decrease $v_y$.

\section{Conclusion}

In conclusion, we study spin-2 condensates with SO coupling.
Due to the competition among SO coupling, spin-dependent,
and spin-independent interactions, the ground states are
found to contain
one or two counter-propagating plane waves respectively in ``ferromagnetic''
and ``polar'' interaction cases, where both the single particle
Hamiltonian and atomic interaction energies can reach their
corresponding minimum values.
For ``cyclic'' interaction case with axisymmetric SO coupling,
the ground states can be categorized into two different types
respectively containing four or three plane waves,
leading to square or triangular patterns correspondingly.
For asymmetric SO coupling, the ground states are determined
by the ratio of two SO coefficients: $|v_y|/|v_x$.
As long as ratio is decreased, the spin component
density distributions can evolve from triangular to square shaped lattices
before asymptotically reaching the striped pattern.
This structure phase transition provides a clear signature for
SO coupling.
In the ``cyclic'' interaction case, the corresponding minimum value
of single particle Hamiltonian and the
interaction energies cannot be reached simultaneously.

\section{Acknowledgments} 

Z.F.X thanks Masahito Ueda for valuable discussions. This work is
supported by NSF of China under Grants No. 11004116 and No.
10974112, NKBRSF of China, and the research program 2010THZO 
of Tsinghua University.


\begin{thebibliography}{10}
\bibitem{fetter2009}
  Alexander L. Fetter, Rev. Mod. Phys. \textbf{81}, 647 (2009).

\bibitem{cooper2008}
  N. R. Cooper, Adv. Phys. \textbf{57}, 539 (2008).

\bibitem{dalibard2010}
  Jean Dalibard, Fabrice Gerbier, Gediminas Juzeli\=unas, and Patrik \"Ohberg,
  arXiv: 1008.5378.

\bibitem{juzeliunas2005}
  G. Juzeli\=unas, P. \"Ohberg, J. Ruseckas, and A. Klein, Phys. Rev. A \textbf{71}, 053614 (2005);
  G. Juzeli\=unas, J. Ruseckas, P. \"Ohberg, and M. Fleischhauer, Phys. Rev. A
  \textbf{73}, 025602 (2006).

\bibitem{gunter2009}
  Kenneth J. G\"unter, Marc Cheneau, Tarik Yefsah, Steffen P. Rath, and Jean Dalibard,
  Phys. Rev. A \textbf{79}, 011604 (2009)

\bibitem{lin2009}
  Y.-J. Lin, R. L. Compton, A. R. Perry, W. D. Phillips, J. V. Porto and I. B. Spielman,
  Phys. Rev. Lett. \textbf{102}, 130401 (2009);
  Y.-J. Lin, R. L. Compton, K. Jimenez-Garcia, J.V. Porto,
  and I. B. Spielman, Nature (London) \textbf{462}, 628 (2009).

\bibitem{ruseckas2005}
  J. Ruseckas, G. Juzeli\=unas, P. \"Ohberg, and M. Fleischhauer,
  Phys. Rev. Lett. \textbf{95}, 010404 (2005);
  Gediminas Juzeli\=unas, Julius Ruseckas, and Jean Dalibard
  Phys. Rev. A \textbf{81}, 053403 (2010).

\bibitem{osterloh2005}
  K. Osterloh, M. Baig, L. Santos, P. Zoller, and M. Lewenstein,
  Phys. Rev. Lett. \textbf{95}, 010403 (2005).

\bibitem{satija2006}
  Indubala I. Satija, Daniel C. Dakin, and Charles W. Clark,
  Phys. Rev. Lett. \textbf{97}, 216401 (2006).

\bibitem{zutic2004}
  I. \v{Z}uti\'c, J. Fabian, S. Das Sarma, Rev. Mod. Phys. \textbf{76}, 323 (2004).

\bibitem{ruostekoski2002}
  Janne Ruostekoski, Gerald V. Dunne, and Juha Javanainen, Phys.
  Rev. Lett. \textbf{88}, 180401 (2002).

\bibitem{stanescu2008}
  Tudor D. Stanescu, Brandon Anderson, and Victor Galitski, Phys. Rev. A
  \textbf{78}, 023616 (2008).

\bibitem{wang2010}
  Chunji Wang, Chao Gao, Chao-Ming Jian, and Hui Zhai, Phys. Rev. Lett.
  \textbf{105}, 160403 (2010).

\bibitem{ho2010}
  Tin-Lun Ho and Shizhong Zhang, arXiv: 1007.0650.

\bibitem{yip2010}
  S.-K. Yip, arXiv: 1008.2263.

\bibitem{lin2011}
  Y.-J. Lin, K. Jim\'enez-Garc\'ia, and I. B. Spielman, Nature (London) \textbf{471}, 83 (2011).

\bibitem{merkl2010}
  M. Merkl, A. Jacob, F. E. Zimmer, P. \"Ohberg, and L. Santos,
  Phys. Rev. Lett. \textbf{104}, 073603 (2010).

\bibitem{braun2010}
  Daniel Braun, Phys. Rev. A \textbf{82}, 013617 (2010).

\bibitem{ueda2010}
  Masahito Ueda and Yuki Kawaguchi, arXiv: 1001.2072.

\bibitem{zhang2005}
  Wenxian Zhang, D. L. Zhou, M.-S. Chang, M. S. Chapman, and L. You,
  Phys. Rev. Lett. \textbf{95}, 180403 (2005);

\bibitem{sadler2006}
  L. E. Sadler, J. M. Higbie, S. R. Leslie, M. Vengalattore, D. M. Stamper-Kurn,
  Nature \textbf{443}, 312 (2006).

\bibitem{saito2009}
  Hiroki Saito, Yuki Kawaguchi, and Masahito Ueda, Phys. Rev. Lett. \textbf{102}, 230403 (2009).

\bibitem{cherng2009}
  R. W. Cherng and E. Demler, Phys. Rev. Lett. \textbf{103}, 185301 (2009).

\bibitem{matuszewski2010}
  Micha\mbox{\l} Matuszewski, Phys. Rev. Lett. \textbf{105}, 020405 (2010).

\bibitem{kronjager2010}
  Jochen Kronj\"ager, Christoph Becker, Parvis Soltan-Panahi, Kai Bongs, and
  Klaus Sengstock, Phys. Rev. Lett. \textbf{105}, 090402 (2010).

\bibitem{scherer2010}
  M. Scherer, B. L\"ucke, G. Gebreyesus, O. Topic, F. Deuretzbacher,
  W. Ertmer, L. Santos, J. J. Arlt, and C. Klempt, Phys. Rev. Lett.
  \textbf{105}, 135302 (2010).


\bibitem{ciobanu2000}
  C. V. Ciobanu, S.-K. Yip, and Tin-Lun Ho, Phys. Rev. A \textbf{61}, 033607 (2000).

\bibitem{koashi2000}
  Masato Koashi and Masahito Ueda, Phys. Rev. Lett. \textbf{84}, 1066 (2000).


\end{thebibliography}
\end{document}